\pdfoutput=1
\documentclass[reprint,aps,prb,floatfix,twocolumn,superscriptaddress,showpacs,
amsmath,amssymb,showpacs]{revtex4-1}
\usepackage{graphicx}
\usepackage{epstopdf}
\usepackage{epsfig}
\usepackage{epsf}
\usepackage{url}
\usepackage[USenglish]{babel}
\usepackage{hyperref}
\def\bcen{\begin{center}}
\def\ecen{\end{center}}
\allowdisplaybreaks
\renewcommand\[{\begin{equation}}
\renewcommand\]{\end{equation}}
\usepackage{verbatim}
\usepackage{xcolor}
\usepackage{natbib}
\usepackage{dblfloatfix}
\usepackage{amsmath, nccmath}
\usepackage[export]{adjustbox}
\usepackage{dcolumn}
\usepackage{bm}
\usepackage{bbm}
\usepackage{lipsum}

\begin{document}
\title{Screening from $e_g$ states and antiferromagnetic correlations in $d^{(1,2,3)}$ perovskites: \\ A $GW$+EDMFT investigation}
\author{Francesco Petocchi}
\affiliation{Department of Physics, University of Fribourg, 1700 Fribourg, 
Switzerland}
\author{Fredrik Nilsson}
\affiliation{Department of Physics, Division of Mathematical Physics, Lund 
University, Professorsgatan 1, 223 63 Lund, Sweden}
\author{Ferdi Aryasetiawan}
\affiliation{Department of Physics, Division of Mathematical Physics, Lund 
University, Professorsgatan 1, 223 63 Lund, Sweden}
\author{Philipp Werner}
\affiliation{Department of Physics, University of Fribourg, 1700 Fribourg, 
Switzerland}
\begin{abstract}
We perform a systematic {\it ab initio} study of the electronic structure of 
Sr(V,Mo,Mn)O$_3$ perovskites, using the parameter-free $GW$+EDMFT method.  This 
approach self-consistently calculates effective interaction parameters, taking 
into account screening effects due to nonlocal charge fluctuations. Comparing 
the results of a 3-band ($t_{2g}$) description to those of a  5-band 
($t_{2g}$+$e_g$) model, it is shown that the $e_g$ states have little
effect on the low-energy properties and the plasmonic features for the first two compounds 
but play a more active role in SrMnO$_3$. In the case of 
SrMnO$_3$ paramagnetic $GW$+EDMFT yields a metallic low-temperature solution on 
the verge of a Mott transition, while antiferromagnetic $GW$+EDMFT produces an 
insulating solution with the correct gap size. We discuss the possible 
implications of this result for the nature of the insulating state above the 
N\'eel temperature, and the reliability of the $GW$+EDMFT scheme. 
\end{abstract}
\date{\today}

\maketitle
%
%
%
%
\section{Introduction}
Transition-metal oxides represent an important test ground for new theoretical 
and computational schemes aimed at a quantitative description of 
electron-electron correlations. In this class of compounds, methods based on a 
single-particle description of the solid do not provide a satisfactory description due 
to the strong many-body interactions within the partially filled and
narrow $d$ or $f$ bands. Emergent properties, 
such as high-temperature superconductivity and other electronic ordering 
phenomena, are the result of subtle competitions between different 
interactions and require an accurate estimation of free energies. Even in the 
absence of symmetry breaking, basic properties of the solids, such as the metallic or 
insulating nature of the ground state, cannot be easily predicted.\cite{Imada1998} 
Theoretical models capturing the essential physics need to be developed, solved, 
and the results compared to experiments. 

A widely used approach is the combination of 
Density Functional Theory (DFT) and Dynamical Mean-Field Theory 
(DMFT).\cite{Georges1996,Anisimov1991} This scheme reduces the problem to a 
multi-orbital Hubbard model with hopping parameters derived using Wannier basis functions.
In many of these calculations the value of the local Coulomb
interaction is chosen to reproduce experimentally observed properties, such as mass 
enhancements or positions of Hubbard bands. With suitably chosen parameters, many 
properties of correlated materials can be described by DFT+DMFT.\cite{Kotliar2004} 
What is not captured by this approach are 
collective long-range charge fluctuations and the resulting dynamical screening effects. 
These dynamical long-range correlations are on the other hand well described by the $GW$ 
approximation,\cite{Hedin1965} one of the most successful methods for the study
of excited-state properties of weakly correlated compounds, such as 
semi-conductors. \cite{Schilfgaarde2006} After more than a decade of development, 
\cite{Biermann03FirstPrinciples,Tomczak12Combined,Ayral2013,Tomczak14Asymmetry,
Huang2014,Boehnke16When} the power of these two methods has been combined into 
a multitier $GW$+EDMFT formalism, which is applicable to moderately and strongly correlated 
materials.\cite{Boehnke16When,Nilsson17Multitier} EDMFT is the extended version of 
DMFT, \cite{Sun2002} which allows to treat the effect of long-range interactions. 
Multitier refers to the fact that different degrees of freedom are treated with different 
physically motivated approximations: the highest energy bands with single-shot 
$GW$, an intermediate energy window within self-consistent $GW$ and only the 
most strongly correlated bands near the Fermi energy within $GW$+EDMFT. This 
separation makes the scheme computationally feasible and can be implemented 
without any double countings. Apart from the choice of these energy windows, 
multitier $GW$+EDMFT is free of adjustable parameters, and thus a true {\it ab 
initio} method. 

To assess the accuracy and predictive power of this approach, it is important to 
test it on a broad range of compounds, and with different choices of energy 
windows. Here, we continue the effort started in 
Refs.~\onlinecite{Boehnke16When,Nilsson17Multitier} and present a systematic 
study of three prototypical perovskite compounds, namely SrVO$_3$, SrMoO$_3$, 
and SrMnO$_3$. These materials exhibit different fillings of the $t_{2g}$ 
orbitals, and hence different correlation effects. SrVO$_3$ and SrMoO$_3$ are 
paramagnetic metals, while SrMnO$_3$ is an antiferromagnetic insulator. 
DFT+DMFT based modeling of these materials often only considered the $t_{2g}$ 
shell,\cite{Pavarini04Mott, nekrasov2006,Lechermann06Dynamical,Backes16Hubbard} 
but here, we also explore the effect of including the almost empty $e_g$ states. This 
provides a consistency check for the multitier scheme, since the three-band and five-band 
treatments should produce consistent results for the low-energy electronic structure. 

As far as the methodology is concerned, detailed descriptions can be found in 
Ref.~\onlinecite{Nilsson17Multitier}. One extension in the present work is the 
implementation of a self-consistency loop with two sublattices, which allows 
to stabilize solutions with antiferromagnetic order.   

The manuscript is organized as follows: in Sec.~\ref{sec_method} we briefly 
summarize the method, the steps of the self-consistency loop and the rationale 
behind the multitier subdivision of the orbital space. In Sec.~\ref{sec_results} 
we compare the results of the five- and three-band models for SrVO$_3$, SrMoO$_3$ and 
SrMnO$_3$. In Sec.~\ref{sec_conclusions} we present our conclusions.
%
%
%
%
\section{Method}\label{sec_method}
In this Section we give a brief overview of the multitier $GW$+EDMFT method 
developed in Refs.~\onlinecite{Boehnke16When,Nilsson17Multitier} and explain its 
extension to antiferromagnetically ordered systems. 
\subsection{$GW$+EDMFT}
By defining a set of localized wave functions $w_{i\mathbf{R}}(\mathbf{r})$, 
where $i$ is an orbital index and $\mathbf{R}$ is a site index, the self-energy 
$\Sigma$ and polarization $\Pi$ can be divided into the local (onsite) 
components $\Sigma^{\mathrm{loc}}$, $\Pi^{\mathrm{loc}}$ and the remaining 
nonlocal components,
\begin{align}
    \Sigma_{i,j}(\mathbf{k},i\nu)= \Sigma^{\mathrm{loc}}_{i,j}(i\nu) + 
\Sigma^{\mathrm{nonloc}}_{i,j}(\mathbf{k},i\nu), \\
    \Pi_{\alpha,\beta}(\mathbf{k},i\omega)= 
\Pi^{\mathrm{loc}}_{\alpha,\beta}(i\omega) + 
\Pi^{\mathrm{nonloc}}_{\alpha,\beta}(\mathbf{k},i\omega).
\end{align}
The Greek indices $\alpha,\beta$ denote a product basis, $\alpha = \{i,j\}$, 
necessary to expand the two-particle functions, and we have assumed that two 
basis functions localized on different sites do not overlap. The self-energy and 
polarization are related to the Green's function $G$ and screened interaction 
$W$ through the Dyson equations
\begin{align}
    G&=G^0 + G^0 \Sigma G, \\
    W&=v + v \Pi W,
\end{align}
where $G^0$ is the bare propagator and $v$ the bare Coulomb interaction. The key 
approximation in EDMFT is that the nonlocal components of $\Sigma$ and $\Pi$ 
are negligible, $\Sigma=\Sigma^{\mathrm{loc}}$ and $\Pi=\Pi^{\mathrm{loc}}$. 
With these approximations the full lattice problem can be mapped to an effective 
local impurity problem with a dynamical bare propagator $\mathcal{G}(i\nu)$ and 
a dynamical bare impurity interaction $\mathcal{U}(i\omega)$. These so-called 
Weiss-fields are determined self-consistently such that the impurity Green's 
function reproduces the local lattice Green's function,  
$G^{\mathrm{imp}}=G^{\mathrm{loc}}$, and correspondingly for the screened 
interaction, $W^{\mathrm{imp}}=W^{\mathrm{loc}}$.

$GW$+EDMFT can be regarded as an extension of EDMFT, where the nonlocal 
components are accounted for within the $GW$ approximation,
\begin{align}
    \Sigma_{ik}^{\mathrm{nonloc}}(\mathbf{q},\tau) = &-\sum_{\mathbf{k}jl} 
G_{jl}(\mathbf{k},\tau) W_{ijkl}(\mathbf{q}-\mathbf{k},\tau) \nonumber \\
    &+ \sum_{jl} G_{jl}^{\mathrm{loc}}(\tau) W_{ijkl}^{\mathrm{loc}}(\tau), 
\label{eq:GWSigma} \\
\Pi^{\mathrm{nonloc}}_{mm'nn'}(\mathbf{q},\tau) = 
&\sum_{\mathbf{k}}G_{mn}(\mathbf{k},\tau)G_{n'm'}(\mathbf{k}-\mathbf{q},-\tau) 
\nonumber \\
&-G^{\mathrm{loc}}_{mn}(\tau)G^{\mathrm{loc}}_{n'm'}(-\tau). \label{eq:GWPol} 
\end{align}
The matrix elements of the screened interaction are defined as
\begin{align}
  \label{eqn:downfolded}
  W_{ijkl}(\mathbf{q},i\omega)=&\sum_{\mathbf{R},\mathbf{R'}} 
e^{i\mathbf{q}(\mathbf{R}-\mathbf{R}')}\int 
d\mathbf{r}d\mathbf{r}'w_{i\mathbf{R}}^{*}(\mathbf{r})w_{j\mathbf{R}}^{}(\mathbf
{r})\nonumber\\
  &\times W(\mathbf{r},\mathbf{r}',i\omega) 
w_{k\mathbf{R}'}(\mathbf{r}')w^{*}_{l\mathbf{R}'}(\mathbf{r}'),
\end{align}
where we once again have assumed that two basis functions localized on different 
sites have zero overlap.

The $GW$+EDMFT self-consistency cycle contains the following steps:
\begin{enumerate}
\item Start with an initial guess for $\Sigma^\mathrm{imp}$, $\Pi^\mathrm{imp}$ 
and $G_\mathbf{k}$.
$\Sigma^\mathrm{loc}=\Sigma^\mathrm{imp}$ and 
$\Pi^\mathrm{loc}=\Pi^\mathrm{imp}$ (EDMFT approximations). \label{nbr:EDMFTcondition}
\item Compute  $\Sigma^{\mathrm{nonloc}}$ and $\Pi^{\mathrm{nonloc}}$ according 
to equations (\ref{eq:GWSigma})-(\ref{eq:GWPol}). \label{nbr:nonloc}
\item Define $\Sigma_\mathbf{k}=\Sigma^\mathrm{imp} + 
\Sigma^\mathrm{nonloc}_\mathbf{k}$ and $\Pi_\mathbf{q}=\Pi^\mathrm{imp} + 
\Pi^\mathrm{nonloc}_\mathbf{q}$ ($GW$+EDMFT approximations). 
\label{nbr:EDMFTcondition}
\item Calculate 
$G_\mathbf{k}=\big((G^{(0)}_\mathbf{k})^{-1}-\Sigma_\mathbf{k}\big)^{-1}$ and 
$W_\mathbf{q}=v_\mathbf{q}\left(\mathbbm{1}-\Pi_\mathbf{q} 
v_\mathbf{q}\right)^{-1}$.
\item Using $G^\mathrm{loc}= \frac{1}{N}\sum_\mathbf{k} G_\mathbf{k}$ and 
$W^\mathrm{loc}=\frac{1}{N} \sum_\mathbf{q}W_\mathbf{q}$ calculate the fermionic 
Weiss field 
\begin{align}
\mathcal{G}=\left(\Sigma^\mathrm{imp}+(G^\mathrm{loc})^{-1}\right)^{-1}
\label{eq:weissfields1}
\end{align}
and the effective impurity interaction 
\begin{align}
\mathcal{U}=W^\mathrm{loc}\left(\mathbbm{1}+\Pi^\mathrm{imp} 
W^\mathrm{loc}\right)^{-1}.
\label{eq:weissfields2}
\end{align}\label{nbr:weissfields}
\item Numerically solve the impurity problem to obtain $G^\mathrm{imp}$ and the 
impurity charge susceptibility $\chi^\mathrm{imp}=\left\langle\hat{n}(\tau)\hat{n}(0)\right\rangle$.
\item Use the current $\mathcal{G}$ and $\mathcal{U}$ to calculate 
$\Sigma^\mathrm{imp}=\mathcal{G}^{-1}-(G^\mathrm{imp})^{-1}$, 
$\Pi^\mathrm{imp}=\chi^\mathrm{imp}\left(\mathcal{U}\chi^\mathrm{imp}-\mathbbm{1
}\right)^{-1}$ and 
$W^\mathrm{imp}=\mathcal{U}-\mathcal{U}\chi^\mathrm{imp}\mathcal{U}$.
\item If the selfconsistency conditions $G^\mathrm{imp} = G^\mathrm{loc}$ and 
$W^\mathrm{imp}= W^\mathrm{loc}$ are not fulfilled within a given tolerance, go 
back to step \ref{nbr:nonloc}.
\end{enumerate}
\subsection{Multitier-approach}
If the self-consistency cycle is performed in the complete Hilbert space the 
$GW$+EDMFT formalism is derivable from a free energy functional $\Psi$ and is 
hence conserving.\cite{Biermann03FirstPrinciples} However, in practice this is 
not feasible. To overcome this problem a multitier implementation was developed 
in Refs.~\onlinecite{Boehnke16When, Nilsson17Multitier}. In this approach the 
complete Hilbert space is divided into three different subspaces, each treated 
at a different level of approximation. Correspondingly the calculations are 
divided into three tiers which refer to the different subspaces. The multitier 
approach is a systematic downfolding procedure from the complete Hilbert space 
to a smaller subspace and includes well-defined double counting corrections at 
each step:

\begin{enumerate}
 \item TIER III: First, a DFT calculation is 
performed using the FLAPW code FLEUR\cite{fleur}. Based on the DFT bandstructure 
we compute the one-shot $GW$ self-energy $\Sigma^{G^{\, 0}W\, ^0}$ using the 
SPEX code.\cite{Friedrich10Efficient,fleur} Then, an intermediate- or low-energy 
subspace, $I$, which includes up to 10 bands around the Fermi energy is defined  
using maximally localized Wannier functions as implemented in the Wannier90 
library.\cite{Marzari97Maximally,Mostofi08wannier90,Freimuth08Maximally,
Sakuma13Symmetryadapted} The effective Coulomb interaction, $U$, on the 
intermediate subspace is computed within the constrained random-phase 
approximation (cRPA) \cite{Aryasetiawan04Frequencydependent} using the SPEX 
code. The $G^{\, 0}W\, ^0$ self-energy contribution from within the intermediate 
subspace is removed from $\Sigma^{G^{\, 0}W\, ^0}$ to define an effective 
bare propagator $G_{\mathbf{k}}^{\, 0}$ for the intermediate subspace. 
 \item TIER II: In the intermediate subspace the self-energy is calculated using 
a custom self-consistent $GW$-implementation (See 
Ref.~\onlinecite{Nilsson17Multitier}). A correlated subspace $C$, which can be 
smaller or equal to the intermediate subspace, is defined. The local part of the 
$GW$ self-energy and polarization from within the correlated subspace is 
subtracted to define the effective bare propagator and effective bare interaction for $C$.
 \item TIER I: At each step of the self-consistency cycle local corrections to 
the self-energy and polarization in the correlated subspace $C$ are computed using EDMFT. 
The effective impurity problem is solved using the CT-Hyb~\cite{Werner06ContinuousTime,Werner07Efficient,Gull11Continuoustime} 
quantum Monte-Carlo algorithm implemented in ALPS,\cite{Bauer11Alps,Alps,Hafermann13Efficient} 
while the self-consistency equations make use of the TRIQS framework.\cite{Parcollet15Triqs} 
The EDMFT calculation provides local corrections to the self-energy and 
polarization within the correlated subspace. 
\end{enumerate}

The complete expressions for the Green's function and screened interactions are:
\begin{align}
&G_{\mathbf{k}}^{-1}=\overbrace{\underbrace{\mathrm{i}\omega _{n}+\mu 
-\varepsilon 
_{\mathbf{k}}^{\mathrm{DFT}}+V_{\mathrm{XC},\mathbf{k}}}_{G_{\mathrm{Hartree},
\mathbf{k}}^{\, 0}{}^{-1}}\underbrace{-\left(\Sigma_{\mathbf{k}}^{G^{\, 0}W^{\, 
0}}-\Sigma_{\mathbf{k}}^{G^{\, 0}W^{\, 0}}\big|_{I}\right)}_{-\Sigma 
_{\mathrm{r},\mathbf{k}}}}^{\text{TIER III},\; G_{I, \mathbf{k}}^{\, 
0}{}^{-1}}\notag\\
&\underbrace{-\left(\Sigma _{\mathbf{k}}^{GW}\big|_{I}-\Sigma 
^{GW}\big|_{C,\mathrm{loc}} + \Delta V_H|_{I} \right)}_{\text{TIER 
II}}\underbrace{-\Sigma ^{\mathrm{EDMFT}}\big|_{C,\mathrm{loc}}}_{\text{TIER 
I}}\;,\label{eqn:fullG} \\
&W_\mathbf{q}^{-1}=\overbrace{v_\mathbf{q}^{-1}\underbrace{-\left(\Pi^{G^{\, 
0}G^{\, 0}}_\mathbf{q}-\Pi^{G^{\, 0}G^{\, 
0}}_\mathbf{q}\big|_I\right)}_{-\Pi_{\mathrm{r},\mathbf{q}}}}^{\text{TIER 
III},\; U_{I,\mathbf{q}}^{-1}}\notag\\
&\underbrace{-\left(\Pi_\mathbf{q}^{GG}\big|_I-\Pi^{GG}\big|_{C,\mathrm{loc}}
\right)}_{\text{TIER 
II}}\underbrace{-\Pi^\mathrm{EDMFT}\big|_{C,\mathrm{loc}}}_{\text{TIER 
I}}.\label{eqn:fullW}
\end{align}
The self-energies is Eq.~(\ref{eqn:fullG}) only contain the exchange and 
correlation parts, while $\Delta V_H|_{I}$ represents the change of the Hartree potential 
within the intermediate subspace (see Ref.~\onlinecite{Nilsson17Multitier} for a 
detailed description). $V_{\mathrm{XC},\mathbf{k}}$ is the exchange-correlation 
potential from the DFT calculation. The notation $A\big|_{I}$ means that all 
internal sums when evaluating $A$ are restricted to the subspace $I$.
\subsection{Antiferromagnetic extension}\label{sec_AFM}
Strongly correlated multiorbital systems at certain integer fillings tend to 
develop long-range magnetic ordering. In particular, on a bipartite lattice, 
there is a strong tendency to antiferromagnetic order at half-filling. At 
sufficiently low temperature we then expect the appearance of a 
solution with a local spin polarization. A staggered magnetization on a bipartite lattice can be easily treated 
in DMFT \cite{Georges1996} by considering two sublattices $A$ and $B$ and 
imposing the following relation between the self-energies:
\begin{align}
    \label{eq:spinflip}
    \Sigma_{\uparrow}^{A}	=\Sigma_{\downarrow}^{B}, \quad 
\Sigma_{\downarrow}^{A}	=\Sigma_{\uparrow}^{B}. 
\end{align}
This allows to reduce the EDMFT calculation to the solution of a single impurity problem, 
while the unit cell used in the lattice self-consistency has to be doubled. We 
extended the multitier formalism in order to include this kind of long-range spin ordering 
by doubling the unit cell of the $GW$ calculations in TIER III and TIER II, 
which doubles the size of the lattice Green's function $G_\mathbf{k}$ and 
screened interaction $W_\mathbf{q}$. The calculation in TIER III 
(which provides the input for TIER II) is kept paramagnetic, but we allow for 
a  spin symmetry breaking at the EDMFT level in TIER I, which feeds back into TIER II. 
Hence, in TIER I, we introduce spin-dependent self-energies
\begin{align}
    \Sigma ^{\mathrm{EDMFT}}\big|_{C,\mathrm{loc}}\longrightarrow 
\Sigma_{\uparrow,\downarrow} ^{\mathrm{EDMFT}}\big|_{C,\mathrm{loc}}.
\end{align}
We do not need to apply any seed, since the Monte Carlo errors enable a 
symmetry breaking in the self-consistent calculation. The spin-dependent local 
self-energies are then associated with the two sites of the lattice Green's function. 
Since the decoupling of the long-range interaction is in the charge channel, implying 
that the local vertex  $\Pi ^{\mathrm{EDMFT}}\big|_{C,\mathrm{loc}}$ is computed via 
the local charge susceptibility, all the two-particle fields remain symmetric 
with respect to the spin index. 

We will apply this extension only to SrMnO$_3$, which meets the requirements for 
antiferromagnetic order in term of filling and interaction strength, and which 
is experimentally found to be in an antiferromagnetic phase at low temperatures.
%
%
%
%
\section{Results}\label{sec_results}
In the following we will investigate the three perovskite compounds SrVO$_3$, 
SrMoO$_3$ and SrMnO$_3$ using two different low-energy models. In the 
$t_{2g}+e_g$ model the correlated subspace (TIER I) contains all five $d$ 
orbitals, while in the $t_{2g}$ model, it is restricted to the $t_{2g}$ 
sub-shell. The calculations for SrVO$_3$ and SrMoO$_3$ are performed at inverse 
temperature $\beta=10$ eV$^{-1}$, corresponding to $T=1160$ K, while in the case 
of SrMnO$_3$ we use $\beta=40$ eV$^{-1}$ corresponding to $T=290$ K. The latter 
value is close to $T_{\text{N\'eel}}$ $\sim$ 260 K for SrMnO$_3$,\cite{Takeda1974,Saitoh1995,Kang2008,Kim2010}  
whose magnetic moments order antiferromagnetically in all directions (G-type ordering). 

\onecolumngrid
\begin{center}
\begin{figure}[ht]
  \includegraphics[width=0.44\textwidth]{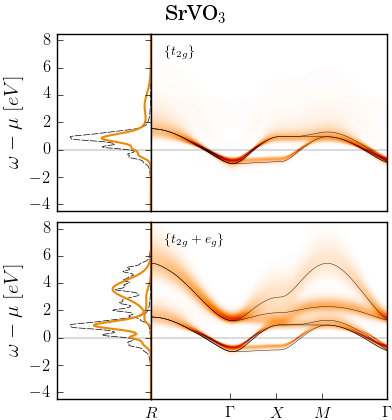} 
  \hspace{5mm}
  \includegraphics[width=0.44\textwidth]{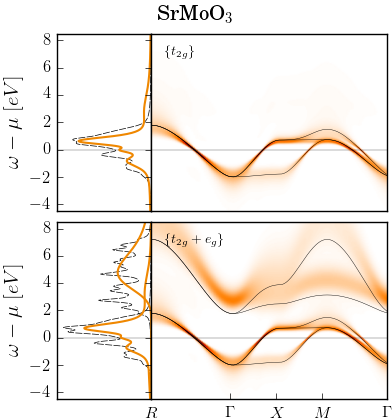}
  \caption{Local and $\mathbf{k}$-resolved spectral function of SrVO$_3$ (left) 
and SrMoO$_3$ (right) in the three- (top) and five- (bottom) band description. Thin black lines represent the LDA bandstructure.\label{SVO_SMoO_Akw} }
\end{figure}
\end{center}
\twocolumngrid

\subsection{SrVO$_3$ and SrMoO$_3$}
SrVO$_3$ is one of the simplest and most extensively studied correlated 
compounds due to its undistorted cubic lattice structure.
\cite{Morikawa95Spectral,Sekiyama04Mutual,Yoshida10Mass,Pavarini04Mott,nekrasov2006,Lechermann06Dynamical,Backes16Hubbard,Tomczak12Combined,Sakuma13Electronic,Tomczak14Asymmetry}
The conduction band is formed by vanadium 3$d$ states of $t_{2g}$ character which are populated by one 
electron per unit cell. Within LDA the conduction band is isolated with a 
bandwidth of roughly 2.2 eV.  The 3.8 eV wide conduction band in SrMoO$_3$ 
originates from the $t_{2g}$ states of the molybdenum cations which are in a $4d^2$ 
configuration. In both systems the DFT calculation predicts empty $e_g$ states 
which start at about 1 eV above the Fermi level. The main difference between the 
two perovskites is thus the filling and the width of the $t_{2g}$ band. 
The experimental photo-emission (PES) and inverse photo-emission (IPES) spectra 
of SrVO$_3$ display a renormalized quasi-particle peak, corresponding to an 
effective mass enhancement of approximately 2, a pronounced upper satellite feature 
at roughly 3 eV and a very weak lower satellite feature at around $-1.5$ eV.\cite{Morikawa95Spectral,Sekiyama04Mutual,Yoshida10Mass,Backes16Hubbard}  
SrMoO$_3$, on the other hand, exhibits a very weakly renormalized quasi-particle 
peak and a pronounced shoulder structure in the PES.\cite{Wadati14Photo} The 
satellite features in the local spectral function of both systems clearly 
indicate correlation effects beyond the LDA. For SrMoO$_3$ it was shown in 
Ref.~\onlinecite{Wadati14Photo} that the satellite features cannot be described 
as Hubbard bands. A later publication, using the same $GW$+EDMFT multitier 
formalism as employed in the current paper, showed that the satellites 
in this compound should instead be interpreted as plasmon satellites originating 
from long-range charge fluctuations.\cite{Nilsson17Multitier} This is 
consistent with the conclusions of Ref.~\onlinecite{Wadati14Photo}, and hence 
there is a relative consensus in the literature on the nature of the satellite 
features in the spectral function of SrMoO$_3$. 
\begin{figure}
  \includegraphics[width=0.48\textwidth]{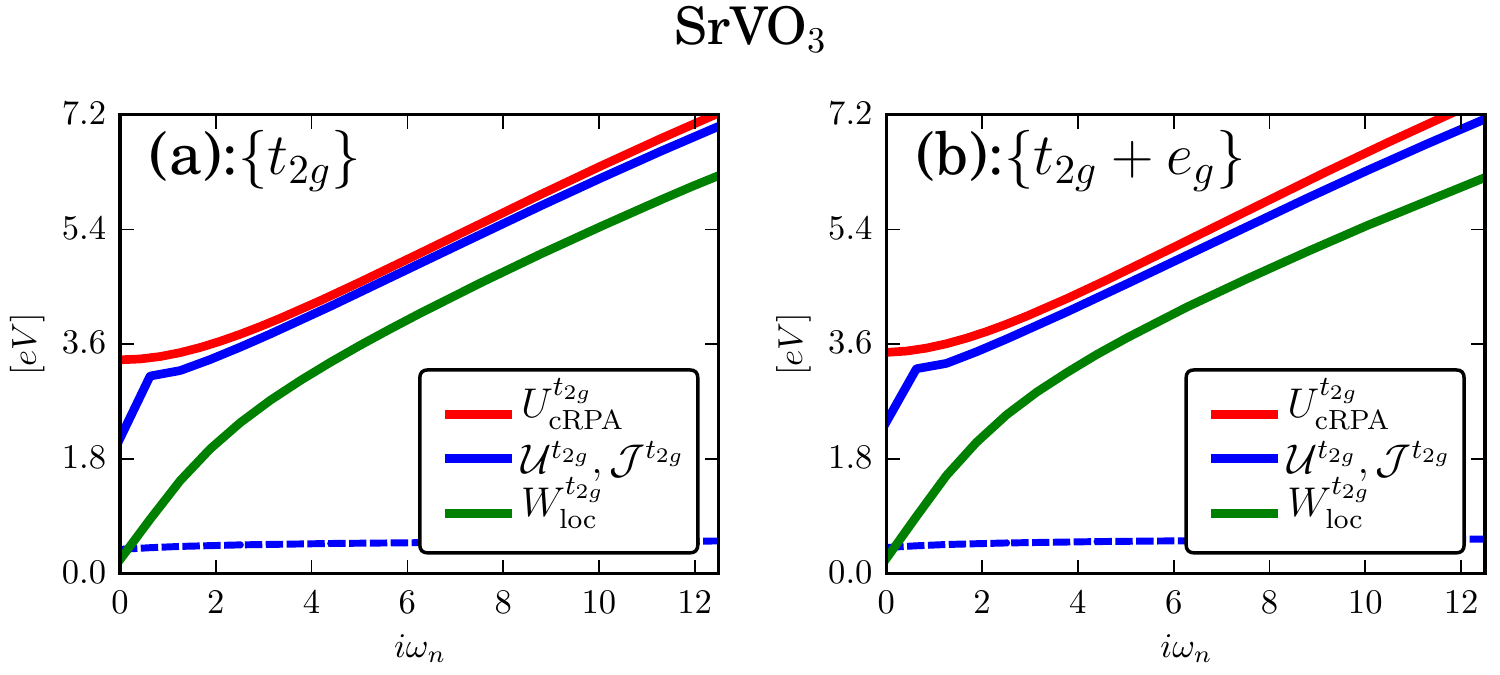}
  \includegraphics[width=0.48\textwidth]{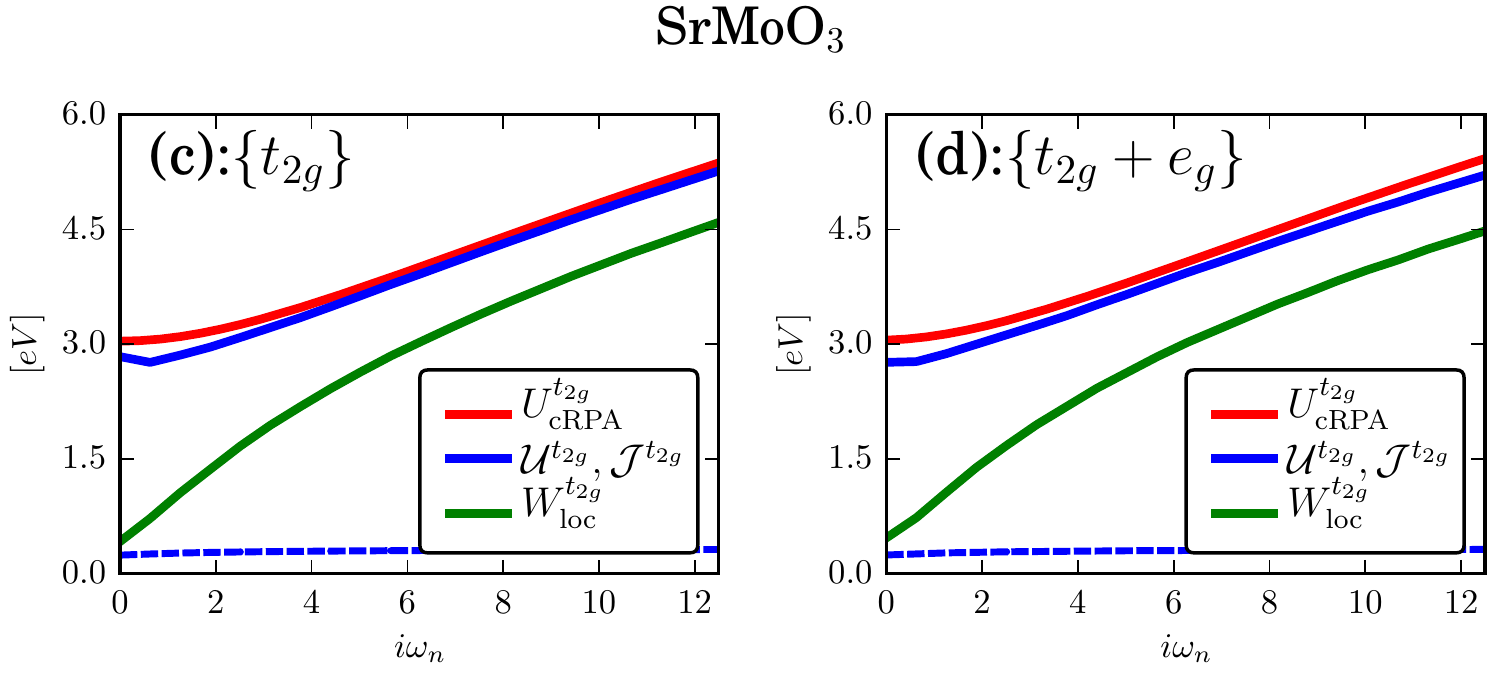}
  \caption{Frequency-dependent cRPA interaction 
$U_\text{cRPA}^{t_{2g}}(i\omega_n)$, Hubbard $\mathcal{U}^{t_{2g}}(i\omega_n)$ 
and Hund's $\mathcal{J}^{t_{2g}}(i\omega_n)$ component (dashed) of the local 
effective interaction and screened interactions 
$W_\text{loc}^{t_{2g}}(i\omega_n)$ for SrVO$_3$ (top) and SrMoO$_3$ (bottom) in 
the three- (a,c) and five- (b,d) band description.\label{SVO_SMoO_UW}}
  \label{fig:VMoint}
\end{figure}
This is not the case for SrVO$_3$, where the origin of the satellite 
features is still under debate. For this compound the 3$d$ valence states (and 
hence also the MLWFs constructed from the $t_{2g}$ bands) are relatively 
localized around the $V$ ion. In cRPA this yields an effective Coulomb 
interaction with a static value of 3.4 eV. LDA+DMFT calculations in 
which the value of $U$ was chosen to reproduce the experimental effective mass 
enhancement,\cite{Pavarini04Mott, nekrasov2006,Lechermann06Dynamical, 
Backes16Hubbard} as well as ab-initio one-shot combinations of $GW$ and DMFT 
\cite{Tomczak12Combined,Sakuma13Electronic,Tomczak14Asymmetry} can roughly 
reproduce the band narrowing and the lower satellite, but place the upper 
satellite observed in IPES too close to the Fermi energy. In 
Ref.~\onlinecite{Tomczak14Asymmetry} it was instead speculated that the upper 
satellite could originate from the $e_g$ states. Common to all the 
above mentioned calculations is that the lower satellite in the SrVO$_3$ spectral function was 
interpreted as a Hubbard band, because of the strong local correlations between 
$t_{2g}$ electrons on the same $V$ site, while the upper satellite was either 
interpreted as originating from the $e_g$ states or left unexplained. 

The interpretation of the satellites as Hubbard bands may be related to the fact that 
DFT+DMFT calculations only include local correlations in the solution of the 
low-energy model. On the other hand, the satellite structures of SrVO$_3$ are well 
described by the cumulant expansion,\cite{Gatti13Dynamical} which is an 
expansion of the Green's function that is based on the $GW$-approximation of the 
self-energy.\cite{aryasetiawan1996, guzzo2011} Because the $GW$ method does not 
capture the strong local correlations that give rise to Hubbard bands, these  
satellite features should be interpreted as plasmons.
\begin{figure}
  \includegraphics[width=0.34\textwidth]{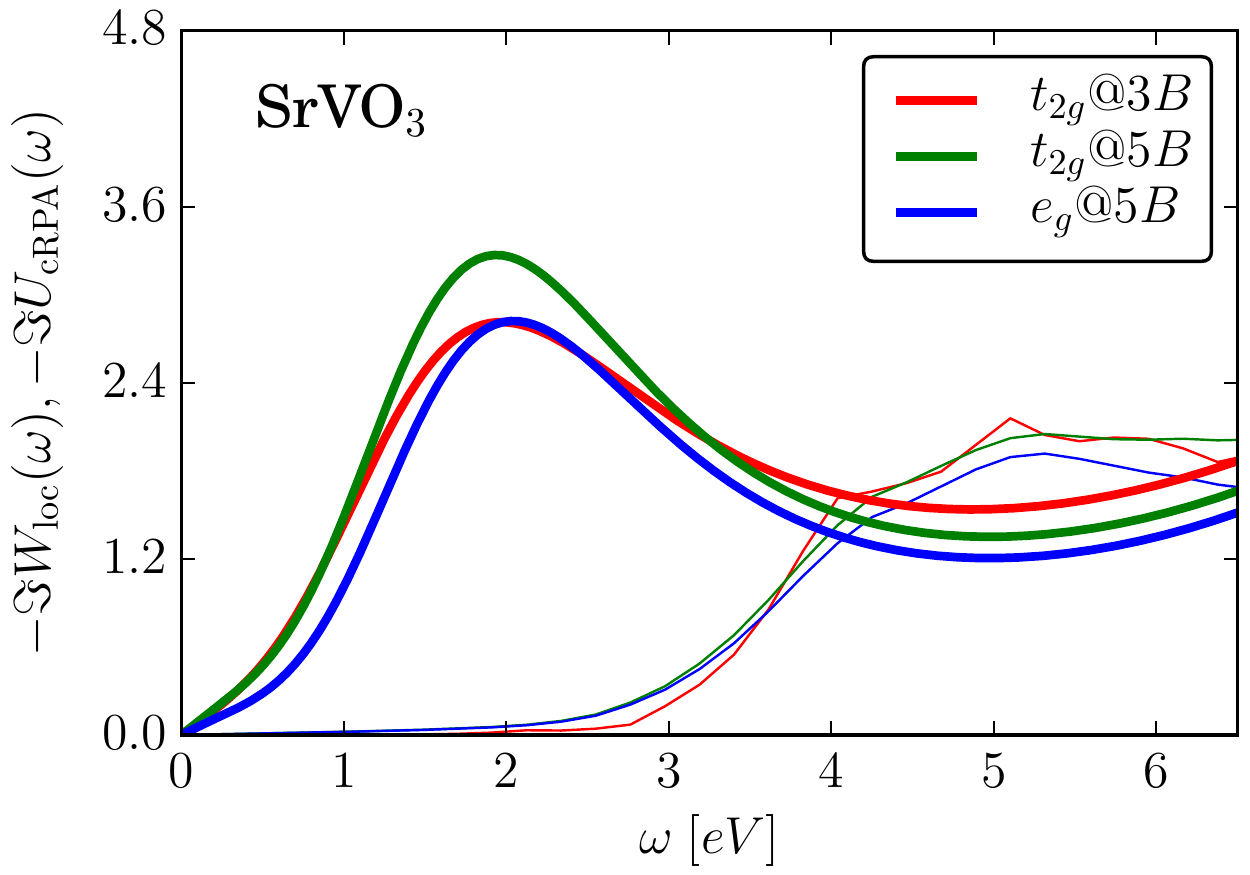}
  
  \vspace{5mm}
  
  \includegraphics[width=0.34\textwidth]{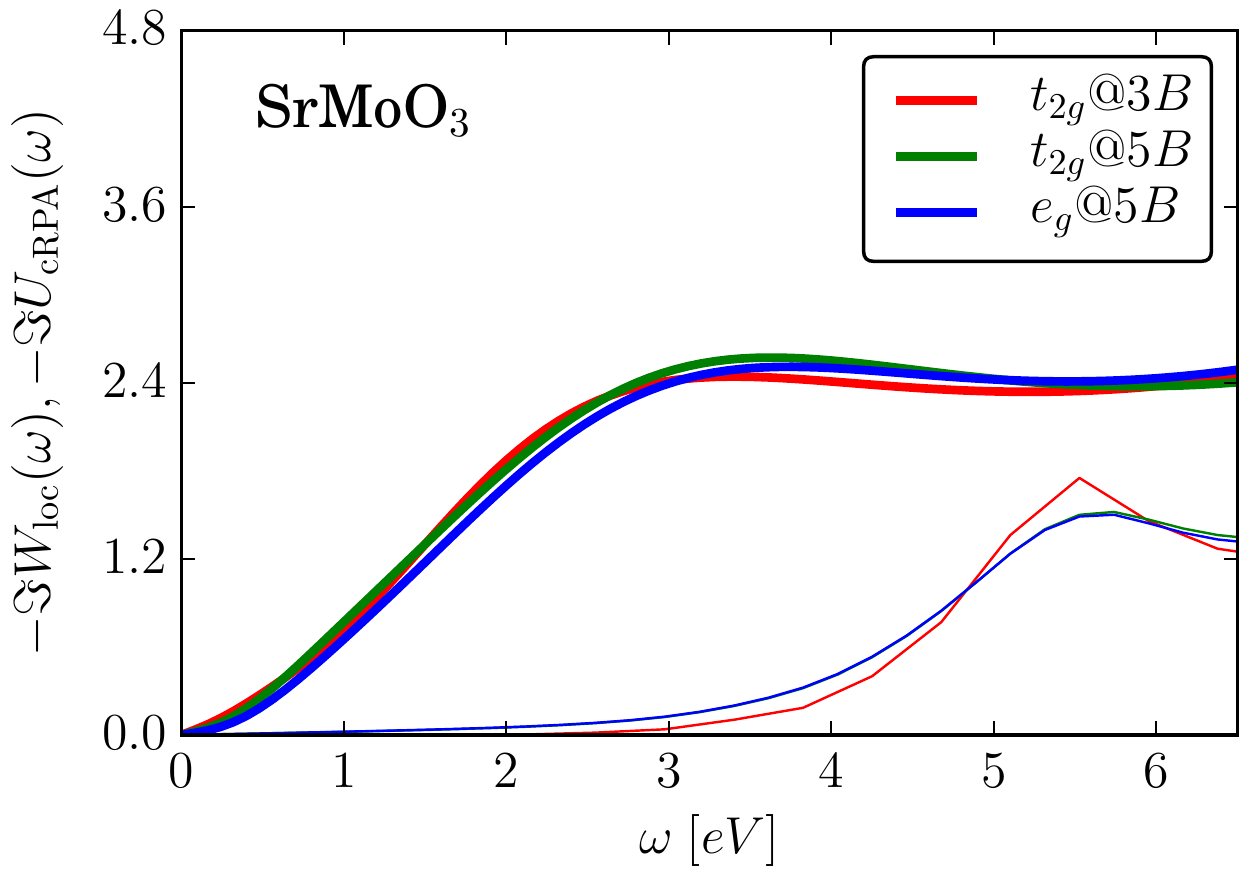}
  \caption{Thick lines represent the imaginary part of fully screened 
interaction on the real frequency axis $-\Im W_\text{loc}^{t_{2g}}(\omega)$ for 
SrVO$_3$ and SrMoO$_3$ while thin lines in the same color code indicate the 
imaginary part of the initial cRPA interaction $-\Im U_\text{cRPA}^{t_{2g}}(\omega)$. 
\label{RealInts}}
\end{figure}

Multitier $GW$+EDMFT considers both types of correlations and therefore 
is a good scheme to investigate the origin of the satellites. In 
Refs.~\onlinecite{Boehnke16When,Nilsson17Multitier} it was shown that the $GW$+EDMFT 
multitier technique yields high energy satellites which are most naturally 
explained in terms of plasmonic excitations when the intermediate and correlated 
subspaces include only the $t_{2g}$ orbitals. This conclusion is supported by 
the relatively small value of the self-consistently computed local interaction 
$\mathcal{U}(0)$, which cannot explain those structures as Hubbard bands, while 
reproducing the experimental mass enhancement relatively well (the band 
renormalization is slightly too small).
In Fig.~\ref{SVO_SMoO_Akw} we show the local and $\mathbf{k}$-resolved spectral 
functions for SrVO$_3$ and SrMoO$_3$ obtained from the three-band and five-band 
calculations. Focusing on the spectral function associated with the $t_{2g}$ 
states, we see that the inclusion of the $e_g$ orbitals has no significant 
effect on the partial $t_{2g}$ spectral function. In particular, the position 
and strength of the satellite features is similar in the three- and five-band models. 
The fact that the satellites at 3 eV in the local spectral 
function follow the dispersion of the unoccupied part of the quasi-particle bands 
is consistent with the plasmon scenario.
We find that the crystal field splitting between the two manifolds is 
significantly enhanced by correlation effects in the case of SrMoO$_3$ while for 
SrVO$_3$ the $e_g$ states remain at the same position as in the LDA 
bandstructure. 

In Fig. \ref{SVO_SMoO_UW} we show the frequency dependent interaction along the 
Matsubara axis. The results of the three- and five-band calculations 
for SrMoO$_3$ do not show any significant difference, probably as a consequence 
of the correlation enhanced crystal field splitting, which decouples the $t_{2g}$ and $e_g$ bands. 
In the case of SrVO$_3$, the screening effects on the local $t_{2g}$ interaction 
$\mathcal{U}^{t_{2g}}(0)$ coming from the inclusion of the $e_g$ orbitals 
are more pronounced. Here the five-band system is characterized by $\mathcal{U}^{t_{2g}}(0)=2.4$ eV 
versus $\mathcal{U}^{t_{2g}}(0)=2.1$ eV in the three-band case. 
This difference is bigger than the difference in the screened cRPA interaction 
provided as input (see red line). The relatively large difference in 
$\mathcal{U}^{t_{2g}}(0)$ may be related to 
the pole at low frequency discussed in Ref.~\onlinecite{Nilsson17Multitier}. A small 
shift of this pole to lower frequencies can lead to substantial changes in the 
static value of the interaction (the results for $\mathcal{U}^{t_{2g}}$ above the 
pole are very similar in both models). In spite of this difference in the local 
interaction strength, long-range charge fluctuations lead to an almost complete 
screening of the interaction, i.e., an almost complete vanishing of $W_\text{loc}^{t_{2g}}(0)$.  

The similarity between the three- and five-band results is also seen in the 
broad pole structure in $-\Im W_\text{loc}(\omega)$ (Fig.~\ref{RealInts}) 
which provides a consistent explanation of the satellites in the spectral 
function in terms of long-range charge fluctuations.
Also, in agreement with our previous studies on the same compounds,\cite{Boehnke16When,Nilsson17Multitier} 
the plasmon peak is higher in SrVO$_3$ indicating 
stronger screening effects compared to SrMoO$_3$.

We conclude from this analysis that the three- and five-band calculations yield 
consistent interpretations of the satellite features in these compounds as 
plasmons rather than Hubbard bands. The presented data also provide a convincing
check for the validity of the downfolding procedure in the multitier approach.
\subsection{SrMnO$_3$}
\subsubsection{Results for paramagnetic SrMnO$_3$}
There is a substantial level of agreement on the importance of electronic 
correlations in SrMnO$_3$, while their role in determining the experimentally observed  
insulating ground state is still debated. Previous studies on this material 
\cite{Mravlje2012,Sondena2006,Bauernfeind2018} employed DFT for the structural 
properties or DFT+DMFT to incorporate the effect of Hubbard-like interactions. A 
common aspect of all these studies has been the {\it ad hoc} choice of the 
Hubbard interaction $U$ and Hund coupling $J_H$, which were chosen with the goal 
of reproducing experimental observations like the band gap or the 
magnetic moment. 

\begin{figure}
  \includegraphics[width=0.48\textwidth]{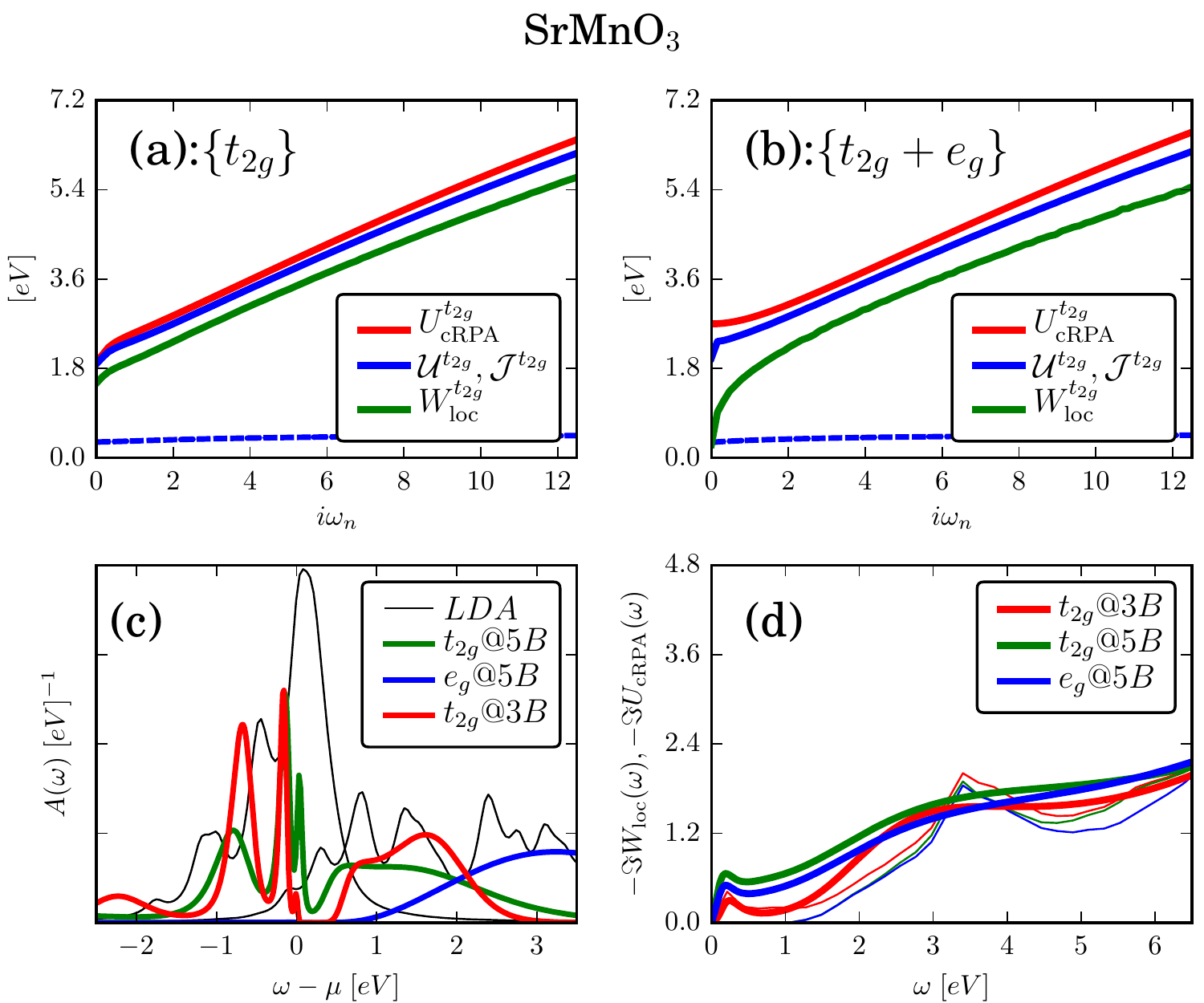}
  \caption{Frequency dependent cRPA interaction $U_{cRPA}^{t_{2g}}(i\omega_n)$, 
Hubbard $\mathcal{U}^{t_{2g}}(i\omega_n)$ and Hund's 
$\mathcal{J}^{t_{2g}}(i\omega_n)$ (dotted blue line) components of the local 
effective interaction and screened interaction 
$W_\text{loc}^{t_{2g}}(i\omega_n)$ for SrMnO$_3$ in the three- (a) and five- (b) 
band case. Panel (c) shows the local spectral function for the different 
correlated subspaces. In (d) the imaginary parts of the cRPA interaction (thin 
lines) and the fully screened (thick lines) interactions are shown on the 
real-frequency axis. \label{SrMnO3Ints}}
\end{figure}

In the following we apply the fully self-consistent and parameter-free 
$GW$+EDMFT approach to the three- and five-band models of SrMnO$_3$ which, at
first glance, appear quite similar to the models described above. 
In SrMnO$_3$, three electrons populate the $t_{2g}$ states, hosted by the manganese 
cation, which results in a 2.5 eV wide band. In contrast to the previous systems, 
already at the DFT level, the $e_g$ manifold crosses the Fermi level, 
which puts a question mark behind the validity of a three-band description for 
this compound. This observation also suggests that the $e_g$ states provide 
nonnegligible screening channels. Indeed, the static values of the cRPA 
interaction reported in Fig.~\ref{SrMnO3Ints} are quite different for the two models, 
namely $U_\text{cRPA}(0)=1.9$ eV in the three-band model and 
$U_\text{cRPA}(0)=2.7$ eV in the five-band model. 
We also notice that, even though the effective local interaction $\mathcal{U}$ 
is similar for the two models, the fully screened interaction 
$W_\text{loc}^{t_{2g}}$ is substantially smaller in the five-band model.
This indicates a more metallic behavior of the five-band model due to 
enhanced screening within the low-energy subspace.

\begin{center}
\begin{figure}[b]
  \includegraphics[width=0.785\textwidth]{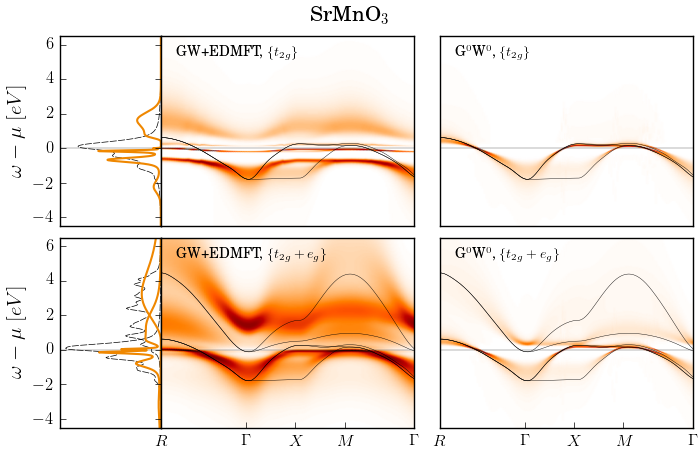}
  \caption{Local and $\mathbf{k}$-resolved spectral function of SrMnO$_3$ obtained using the 
$GW$+EDMFT method (left) and with single-shot $GW$ (right) in the three- (top) 
and five-band (bottom) description. Thin black lines represent the LDA bandstructure.
\label{SrMnO3Akw}}
  \vspace{1cm}
\end{figure}
\end{center}

Similar conclusions can be reached from the lower right panel
of the same figure showing $\Im W_\text{loc}$ along the real frequency axis. $\Im W_\text{loc}^{t_{2g}}$ 
closely follows the cRPA $U$ for low frequencies ($<1$ eV), which indicates that
the metallic screening within the $t_{2g}$ subspace is weak in the three-band model.
In this case the dominant low-energy screening is in the $t_{2g}$-$e_g$ channel
which is incorporated into the cRPA interaction.
On the other hand, in the five-band model the low-energy screening is stronger even though the $e_g$ 
states are pushed up in energy (see panel (c)) and therefore should not contribute 
significantly to screening channels below 1 eV. 
Hence, in the five-band model, the screening within the $t_{2g}$ subspace is enhanced.
This can be understood from the larger weight of the quasiparticle peak in the five-band 
case (see Fig.~\ref{SrMnO3Akw} and the discussion below) which corresponds to an increased 
$t_{2g}$ spectral weight around the Fermi energy.
Again this indicates that an effective model containing only the $t_{2g}$ states might 
be insufficient in describing the screening effects in paramagnetic SrMnO$_3$. 

In Fig.~\ref{SrMnO3Akw} we show the local and 
$\mathbf{k}$-resolved spectral functions of SrMnO$_3$ in the paramagnetic phase 
obtained by the $GW$+EDMFT approach and compare it with the result from a 
single-shot $GW$ calculation. The latter yields a dispersion similar to SrVO$_3$ 
and SrMoO$_3$ (see Ref.~\onlinecite{Nilsson17Multitier}) with a smaller bandwidth and a 
plasmonic broadening occurring mainly in the proximity of the $\Gamma$ and $X$ 
points. The inclusion of local vertex corrections beyond $GW$, however, has 
striking effects on SrMnO$_3$: the near-Mottness of the compound, in both the 
three- and five-band models alike, manifests itself with the formation of broad 
structures centered at $\pm$ 1 eV and with an extremely narrow peak at the Fermi 
level. We identify these features as Hubbard bands considering that their 
separation agrees with the magnitude of the static local effective 
interaction $\mathcal{U}^{t_{2g}}(0)=1.8$ eV. In addition, especially in the 
three-band model, within each of the three main structures (the two Hubbard 
bands at $\sim\pm$ 1 eV and the narrow quasi-particle band) it is possible to 
recognize renormalized and/or broadened replicas of the noninteracting 
dispersion. This behavior is typical of the Mott transition scenario. 
\cite{Georges1996} The asymmetry between the occupied and unoccupied parts of 
the spectra appears to be a consequence of the $GW$-derived 
$\mathbf{k}$-dependent self-energy, which is known to produce such effects.\cite{Casula2016} 
It is also worth noting that these strong correlation effects 
occur even though the ratio between $\mathcal{U}^{t_{2g}}(0)$ and the 
bandwidth is similar to the previous two compounds. This is a Hund coupling 
effect, which leads to a suppression of the kinetic energy at half-filling. As a 
result, the critical interaction for the Mott transition in a multi-orbital 
system with $J_H>0$ is lowest at half-filling.\cite{Werner2009}

\onecolumngrid
\twocolumngrid

The low-energy structures in $\Im W_\text{loc}(\omega)$ will give rise to weak satellites (or broad 
tails) on the high energy side of the Hubbard bands, a feature seen in the local 
$t_{2g}$ spectral function shown in Fig.~\ref{SrMnO3Akw}. Similar physics was investigated for model systems
in Ref.~\onlinecite{Huang2014}, where it was shown that a Mott gap in the fermionic 
spectral function is associated with a peak in $\text{Im}\mathcal{U}(\omega)$ at 
$\omega$ corresponding to the characteristic energy for charge excitations 
across this gap. In the presence of a 
quasi-particle band, there are also screening modes associated with transitions 
between the quasi-particle band and the Hubbard bands. The situation for real materials is, as discussed above, more complicated since
there are multiple screening channels giving rise to different peaks in $\Im W(\omega)$ and a careful analysis of the different screening
channels is needed to clarify the origin of the satellite features.

\begin{figure}
  \includegraphics[width=0.34\textwidth]{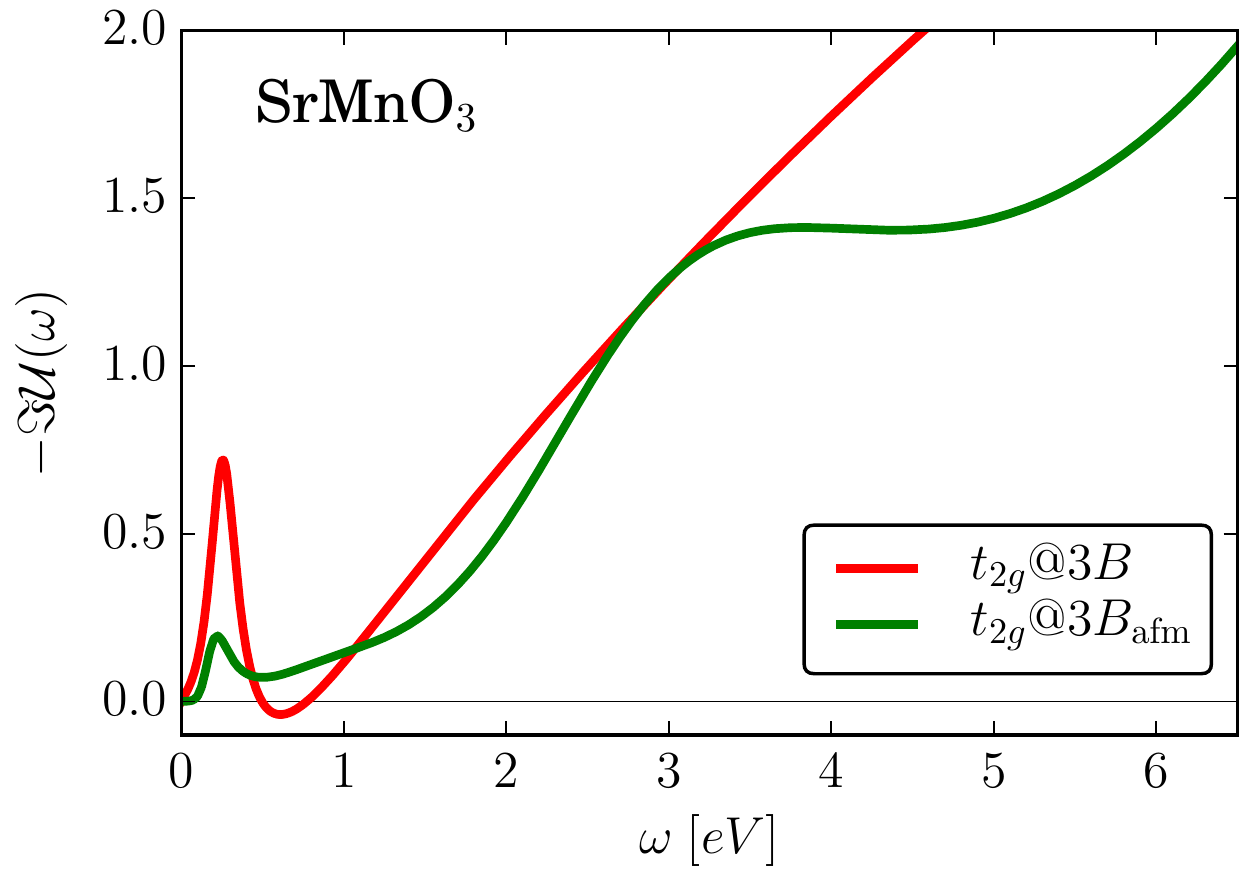}
  \caption{Imaginary part of the local effective interaction on the real 
frequency axis $-\Im \mathcal{U}^{t_{2g}}(\omega)$ for the three-band model of 
SrMnO$_3$ in the paramagnetic (red line) and antiferromagnetic (green line) 
phases. The peaks mainly originate from excitations from outside the $t_{2g}$ 
subspace, which are transferred from $U_\text{cRPA}$, but also include contributions 
from excitations within the quasi-particle band (paramagnetic case) and 
excitations across the gap (antiferromagnetic case), respectively.
\label{curlyUReal}}
\end{figure}

In contrast to the results in previous LDA and LDA+DMFT studies with ad-hoc parameters,\cite{Mravlje2012,Sondena2006}   
both our models of SrMnO$_3$ remain conducting in the paramagnetic phase. The 
metallicity is due to a very narrow quasi-particle band pinned at the chemical 
potential. The quasi-particle peak in the five-band case is a bit larger, see 
Fig.~\ref{SrMnO3Ints}(c), and as a result, the screened interaction 
$W_\text{loc}^{t_{2g}}$ is smaller (panel (d)). Similarly to the case of 
SrMoO$_3$, the $e_g$ center of mass is shifted to higher energies compared to 
LDA. However, a low amount of $e_g$ spectral weight remains in the occupied part 
of the spectral function and is responsible for the self doping effect on the 
$t_{2g}$ states, as one can infer from the peak at the Fermi level not being at 
the center of the gap. The self doping from the $e_g$ states alters the partial $t_{2g}$ 
filling slightly away from half-filling which tends to reduce the local correlations, 
as discussed above. Hence it is this self doping that is responsible for the 
larger quasiparticle peak in the five-band model.
From these observations we conclude that, if restricted to 
the paramagnetic case, the $GW$+EDMFT approach applied to the three- and 
five-band models of SrMnO$_3$  yields a metal on the verge of a Mott transition, in 
which the $e_g$ bands play an active role in determining the overall physics of 
the system.
\subsubsection{Antiferromagnetic phase of SrMnO$_3$}
By construction, the paramagnetic $GW$+EDMFT calculation cannot account for the 
magnetic ground state which is experimentally observed in the cubic phase of 
SrMnO$_3$.\cite{Takeda1974,Saitoh1995,Kang2008,Kim2010} Measurements of the 
magnetic moment report a value of 2.6$\pm$0.2 $\mu_B$ and previous DFT+DMFT 
calculations yield compatible results.\cite{Mravlje2012} The low temperature 
behavior has been reported to be well described by the ordering of $S=3/2$ local 
moments with a $T_{\text{N\'eel}}$  between 233 K and 260 K. The variations in 
the ordering temperature can be accounted for if oxygen defects are considered. 

\begin{figure}
  \includegraphics[width=0.4\textwidth]{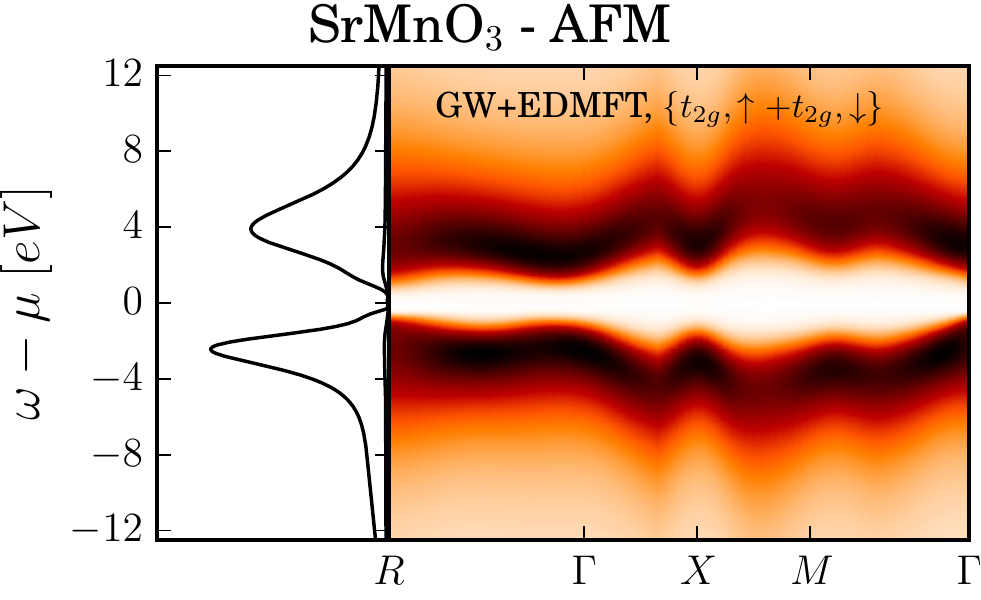}
  \caption{$\mathbf{k}$-resolved spectral function of a three-band model of 
SrMnO$_3$ with long range antiferromagnetic order.\label{SrMnO3Akw_AFM}}
\end{figure}

To describe antiferromagnetic ordering we extended the $GW$+EDMFT multitier 
approach to a bipartite lattice as described in Sec.~\ref{sec_AFM}. At 
$\beta=40$ eV$^{-1}$, the solution with G-type antiferromagnetism 
self-consistently emerges in our parameter-free simulation. The local and 
$\mathbf{k}$-resolved spectral functions of the three-band model are reported in 
Fig.~\ref{SrMnO3Akw_AFM} and exhibit a gap of about 0.5 eV, as well as pronounced 
features at $\pm 2$ eV. The position of the lower satellite is in good agreement with 
the PES measurements of Ref.~\onlinecite{Kang2008,Kim2010}, which were however taken above 
$T_{\text{N\'eel}}$. There are spectral weight tails up to very high energy, 
consistent with plasmonic sidebands. From the imbalance in the spin population 
on a given site we compute the magnetic moment as
\begin{align}
   m=\sum_{\alpha}\left(n_{\alpha\uparrow}-n_{\alpha\downarrow}\right)\mu_{B},
\end{align}
where $\alpha$ is the orbital index. The result is 2.05 $\mu_{B}$ on both 
sublattices. This is in reasonably close agreement with the above-quoted 2.6 
$\mu_{B}$ which has been experimentally found at a much lower temperature.\cite{Takeda1974}
Due to numerical stability issues we could not go below 290 K 
in our calculations, which means that our simulation results are a bit above the 
experimental $T_{\text{N\'eel}}$. However, in mean-field based approaches such 
as DMFT, $T_{\text{N\'eel}}$ is expected to be substantially overestimated since 
long-range fluctuation are neglected. In the $GW$+EDMFT approach used in this 
work, long-range spin fluctuation are not included, as the long range 
interaction is decoupled in the charge channel. We can thus not expect a significant 
improvement in the description of magnetic ordering temperatures. The value of 
$m$ is also reduced compared to the low-temperature value, because our 
simulation temperature is close to $T_{\text{N\'eel}}$. The 
magnetic moment should increase as the temperature is lowered toward the experimental 
value. Due to the lack of adjustable parameters, this kind of 
temperature-dependent analysis would be an interesting topic for future studies.

\begin{figure}
  \includegraphics[width=0.48\textwidth]{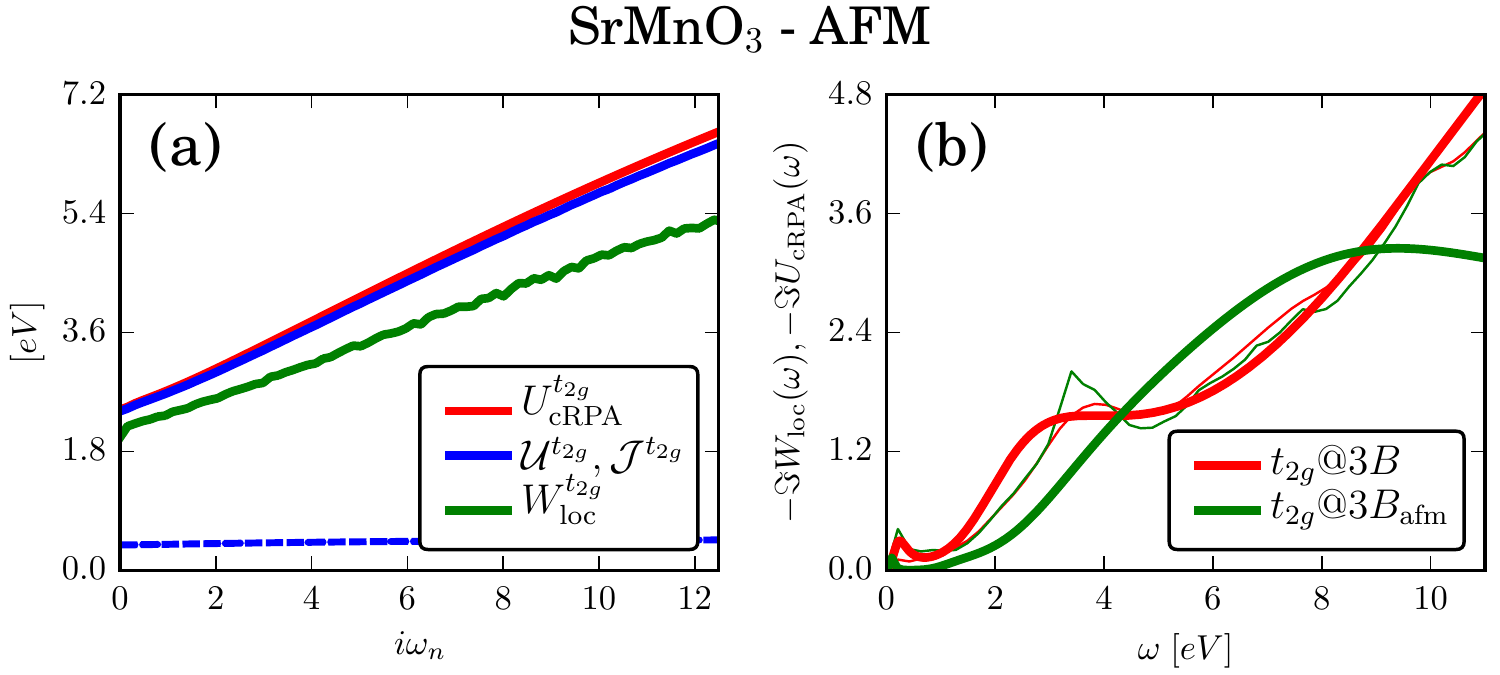}
  \caption{(a) Frequency dependent cRPA interaction 
$U_\text{cRPA}^{t_{2g}}(i\omega_n)$, Hubbard $\mathcal{U}^{t_{2g}}(i\omega_n)$ and 
Hund's $\mathcal{J}^{t_{2g}}(i\omega_n)$ (dotted blue line) components of the 
local effective interaction and screened interaction 
$W_\text{loc}^{t_{2g}}(i\omega_n)$ for the symmetry broken phase of SrMnO$_3$ in 
the three-band model. (b) Imaginary part of the fully screened interaction $-\Im 
W_\text{loc}^{t_{2g}}(\omega)$ (thick lines) and the initial cRPA interaction 
$-\Im U_\text{cRPA}^{t_{2g}}(\omega)$ (thin lines) on the real frequency 
axis. \label{SrMnO_afm_Ints}}
\end{figure}

Because of the heavy numerical cost of the two-sublattice calculation, we 
analyze the symmetry-broken phase only for the three-band model of SrMnO$_3$. 
The insulating nature of the solution should result in small screening effects 
within the correlated subspace. Indeed, as shown in 
Fig.~\ref{SrMnO_afm_Ints}(a), the effective local interaction is essentially 
equal to the cRPA result and the fully screened interaction 
$W_\text{loc}^{t_{2g}}$ is only slightly smaller. The bosonic spectrum $\Im 
W_\text{loc}^{t_{2g}}(\omega)$, shown in panel (b), is also similar to $\Im 
U_\text{cRPA}^{t_{2g}}(\omega)$, but shifted to slightly higher energy (due to the 
gap) and without a prominent feature near 4 eV. The gap in the spectrum 
suppresses the low-frequency screening and introduces a screening channel
corresponding to transitions across the gap which modify the low-energy peak in $\Im W_\text{loc}(\omega)$.
On the other hand, the local effective interaction on the real axis, shown in Fig.~\ref{curlyUReal}, features a broad 
pole centered at $\omega=3$ eV, which is inherited from the corresponding 
peak in $\Im U_\text{cRPA}^{t_{2g}}(\omega)$.
%
%
%
%
\section{Conclusions}\label{sec_conclusions}
We used the recently developed multitier $GW$+EDMFT approach to perform a 
systematic analysis of the electronic properties in a family of transition metal 
perovskites. This self-consistent computational scheme captures 
both local Hubbard physics and long range charge fluctuation, and does 
not rely on any choice of parameters, apart from the energy windows defining the 
model subspaces (tiers). Access to the frequency dependence of the 
selfconsistently determined interactions allows to discriminate spectral 
features originating from local physics and from plasmonic excitations. The 
latter are collective charge fluctuations that screen the local effective 
interaction below a frequency that depends on the details of the correlated 
electronic structure. Both effects are self-consistently accounted for on equal 
footing, making $GW$+EDMFT a fully {\it ab initio} approach. 

The three perovskites considered, namely SrVO$_3$, SrMoO$_3$ and SrMnO$_3$, 
contain 1, 2, and 3 electrons in the $t_{2g}$ shell, respectively, while the 
$e_g$ shell is essentially empty. Due to the effects of filling and Hund 
coupling, the correlations in these three materials are qualitatively different. 
For example, the first two materials are correlated metals, while SrMnO$_3$ is 
an antiferromagnetic insulator. Reproducing these basic properties is a first 
important test for a parameter-free {\it ab initio} scheme. The comparison between the 
three-band and five-band description serves as an additional consistency check 
for the multitier $GW$+EDMFT approach, which should produce consistent results for 
the low-energy electronic structure, independent of the choice of subspace for the self-consistency cycle.  
\begin{table}
\begin{ruledtabular}
\begin{tabular}{ccccccc}
                        &   \multicolumn{2}{c}{SrVO$_3$}   &   
\multicolumn{2}{c}{SrMoO$_3$}  &   \multicolumn{2}{c}{SrMnO$_3$}  \\
                                         &    $t_{2g}$ &   $t_{2g}+e_g$   &     
$t_{2g}$     &   $t_{2g}+e_g$    &      $t_{2g}$     &   $t_{2g}+e_g$   \\ 
\hline
$\mathcal{U}^{t_{2g}}$(0)&      2.13      &      2.38      &      2.78      &    
  2.76      &      1.88       &      2.01       \\
$\mathcal{J}^{t_{2g}}$(0) &      0.38      &      0.41      &      0.24      &   
   0.24      &      0.32       &      0.31      \\
                  $Z^{t_{2g}}$   &      0.62      &      0.62      &      0.7    
    &      0.7        &      0.07       &      0.08      \\
\end{tabular}
\end{ruledtabular}
\caption{\label{tab:tableInteractions}Screened effective interaction parameters 
and quasi-particle weight $Z$ in the $t_{2g}$ sub-shell for the different low 
energy models considered.}
\label{Ztable}
\end{table}
In the case of SrVO$_3$ we obtained a quasi-particle weight in the $t_{2g}$ 
shell which is slightly larger than the  $Z^{t_{2g}}\approx 0.5$ determined in 
photoemission studies,\cite{Yoshida10Mass} see Tab.~\ref{Ztable}. Both in the 
three-band and five-band description, the satellite features appear to have a 
plasmonic origin, since the self-consistently computed static interaction is too 
small to produce Hubbard bands. Hence, within $GW$+EDMFT, SrVO$_3$ is described as 
a correlated metal with strong nonlocal screening effects within the $t_{2g}$ 
subspace. The results for SrMoO$_3$ indicate an even more weakly correlated 
metal. In both these metallic systems the inclusion of the $e_g$ states has 
little effect on the $t_{2g}$ states, which makes the interpretation of the 
satellite structures in terms of plasmonic excitations resilient against the 
choice of the low energy window. 

A qualitatively different situation is encountered in the case of SrMnO$_3$ 
which is experimentally found to be insulating both above 
\cite{Kang2008,Kim2010} and below $T_\text{N\'eel}\approx 290$ K. If restricted 
to paramagnetic solutions, $GW$+EDMFT predicts a strongly correlated metal in 
proximity to a Mott transition, in both the three- and five-band models. The 
Hubbard bands of this strongly correlated metal are at too low energy compared to 
experiment. On the other hand the extension of the method to states with broken 
spin symmetry produces an antiferromagnetic solution with spectral features and 
a magnetic moment in good agreement with experiments. There are two possible 
conclusions one can draw from this observation: 

(i) The (short-range) 
antiferromagnetic spin correlations at $T\approx 300$ K may 
still be so strong that the material is more accurately described by the 
antiferromagnetic solution than by the paramagnetic solution, which ignores 
nonlocal correlations completely. In this case, SrMnO$_3$ would not be a pure 
Mott insulator above $T_{\text{N\'eel}}$, but a material strongly influenced by 
short-range magnetic correlations. These short-range antiferromagnetic fluctuations are, 
in the paramagnetic phase, described by the nonlocal vertex which is not included in the present calculations.
On the other hand, below the N\'eel temperature antiferromagnetic long-range order can be 
incorporated without the need to include the nonlocal vertex, as described in Section IIC.
Hence the lack of a nonlocal vertex is a likely reason for why the 
$GW$+EDMFT method only yields a correct description of SrMnO$_3$ in the antiferromagnetic phase.

(ii) The second possibility is that the interaction parameters which are 
self-consistently computed in $GW$+EDMFT are too small. 
Recent model studies have shown that cRPA can strongly overestimate the 
screening from bands which are relatively close to the Fermi level,\cite{Honerkamp2018} 
which may lead to an underestimation of $U_\text{cRPA}$, and hence of the effective bare interactions in tiers II and I.
However, we have to note that both the five- and three-band models yield metallic solutions,
and that the five-band model actually is more metallic. 
This indicates that the problem with $U_\text{cRPA}$, if it exists at all, is not related to screening from $e_g$ states. 

If the first scenario turns out to be correct, it implies that SrMnO$_3$, 
similar to SrVO$_3$,\cite{Boehnke16When} is a material whose physics has been 
incorrectly described by standard DFT+DMFT treatments. In these calculations, 
interaction parameters are chosen ad-hoc to reproduce spectral features (e. g. 
Hubbard) based on pre-conceived notions about the nature of material. 
The second scenario can be checked by systematically enlarging the low-energy 
space (tier I). 
As we mentioned, the inclusion of the $e_g$ states in SrMnO$_3$ significantly 
alters the fermionic and bosonic Weiss fields. It is then conceivable that 
states outside our correlated subspace, but located at a similar energy 
separation require a treatment beyond $U_\text{cRPA}$. In particular, it would be 
interesting to include the oxygen $p$ orbitals in tier I, since these may 
produce a significant screening. Indeed, in Ref.~\onlinecite{Bauernfeind2018} it 
was shown that these states lie not far from the Fermi level in SrMnO$_3$, while 
the authors of Ref.~\onlinecite{Mossanek2009} even argue that they must be 
included in the low energy models of all the perovskites. 

While the current study showed that the $GW$+EDMFT approach can handle different 
correlations strengths and magnetic phases in material specific setups, and 
produces consistent results for different choices of low-energy models, it will 
be important to perform additional tests on experimentally well-characterized 
compounds. The great strength of $GW$+EDMFT is that it is free from ad-hoc 
parameters and capable of treating weakly and strongly correlated systems. 
However, the treatment of nonlocal correlations is 
limited to charge fluctuations whereas spin fluctuations are not included.
Furthermore, the initial downfolding in tier III relies on cRPA. Additional studies are needed 
to establish for which class of compounds the method has predictive power.

\begin{acknowledgments}
FP and PW acknowledge support from the Swiss National Science Foundation 
through NCCR MARVEL and the European Research Council through ERC Consolidator 
Grant 724103. FN and FA acknowledge financial support from the Swedish Research Council (VR).
The calculations were performed on the Beo04 cluster at the 
University of Fribourg and on resources provided by the Swedish National Infrastructure for Computing (SNIC) at LUNARC.
We thank A.~Georges, M.~Zingl and J.~Mravlje for insightful discussions.
\end{acknowledgments}
\bibliography{paper}
\end{document}